\documentclass[a4paper,11pt]{article}
\usepackage{pos}
\usepackage{subcaption}

\title{A model-independent analysis of neutrino flares detected in IceCube from X-ray selected blazars}
 \ShortTitle{Neutrino flares from X-ray selected blazars}

\author{The IceCube Collaboration \\{\normalsize \normalfont \href{http://icecube.wisc.edu/collaboration/authors/icrc21\_icecube}{http://icecube.wisc.edu/collaboration/authors/icrc21\_icecube} }}

\emailAdd{ankur.sharma@icecube.wisc.edu}
\emailAdd{erin.osullivan@icecube.wisc.edu}

\abstract{Blazars are among the most powerful steady sources in the Universe. Multi-messenger searches for blazars have traditionally focused on their gamma-ray emission, which can be produced simultaneously with neutrinos in photohadronic interactions. However, X-ray data can be equally vital to constrain the SED of these sources, since the hadronically co-produced gamma-rays could get absorbed by the ambient photon fields and cascade down to X-ray energies before escaping. 
In this work, we present the outline for an untriggered, time-dependent analysis of neutrino flares from the direction of X-ray selected blazars using 10 years of IceCube data. A binomial test will be performed on the population to reveal if a subcategory of sources has statistically significant emission. The sources are selected from RomaBZCat, and the p-values and best-fit flare parameters are obtained for each source using the method of unbinned likelihood maximisation.\\

\vspace{4mm}
{\bfseries Corresponding authors:}
Ankur Sharma$^{1*}$, Erin O. Sullivan$^{1}$\\
{$^{1}$ \itshape Dept. of Physics and Astronomy, Uppsala University, Box 516, SE-75120 Uppsala, Sweden}\\
$^*$ Presenter
}

\FullConference{37$^{\rm{th}}$ International Cosmic Ray Conference (ICRC 2021)\\
		July 12th -- 23rd, 2021\\
		Online -- Berlin, Germany}

\begin{document}
\maketitle

\section{Motivation}
 
Blazars are widely favored to be the progenitors of EeV cosmic rays and PeV neutrinos along with TeV gamma-rays, due to the prime conditions in their jets for acceleration of cosmic rays. They have strong magnetic fields that can trap particles, ambient radiation within or proximal to the jets for the particles to interact with, and large scale structure to accommodate acceleration to the highest energies observable.

\begin{figure}[h]
\centering
\begin{subfigure}{.42\textwidth}
  \centering
  \includegraphics[width=\linewidth]{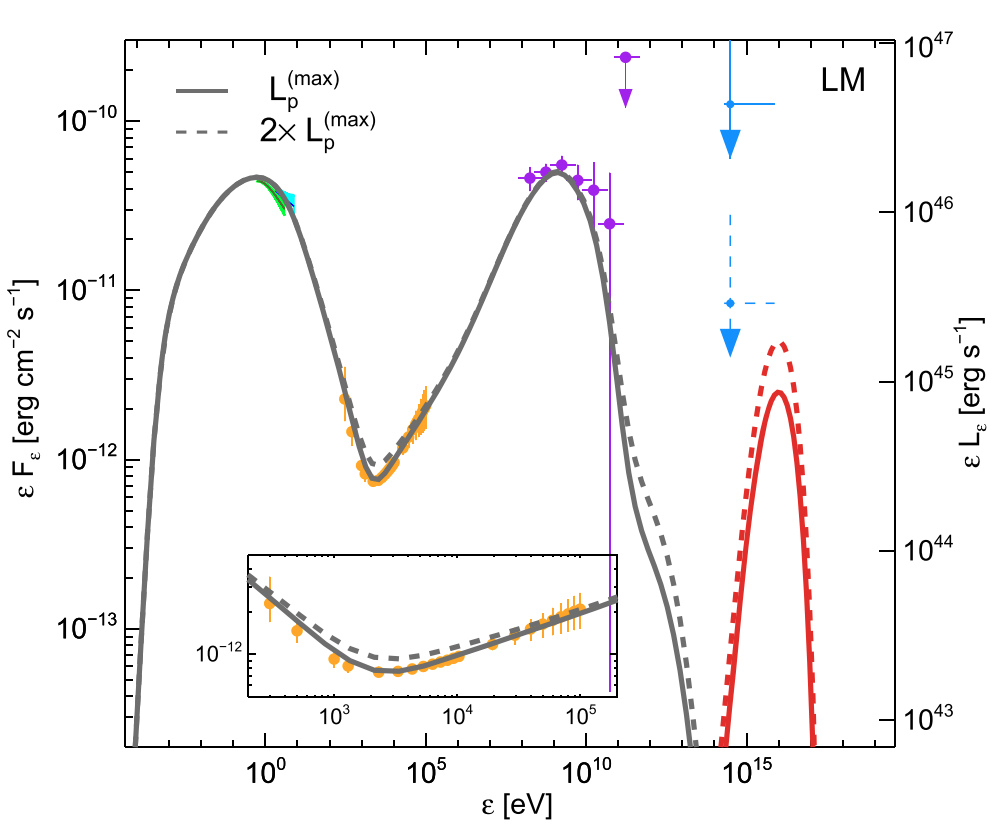}
  \caption{}
  \label{fig:txs_xray_keivani}
\end{subfigure}%
\hspace{1cm}
 \begin{subfigure}{.4\textwidth}
  \centering
  \includegraphics[width=\linewidth]{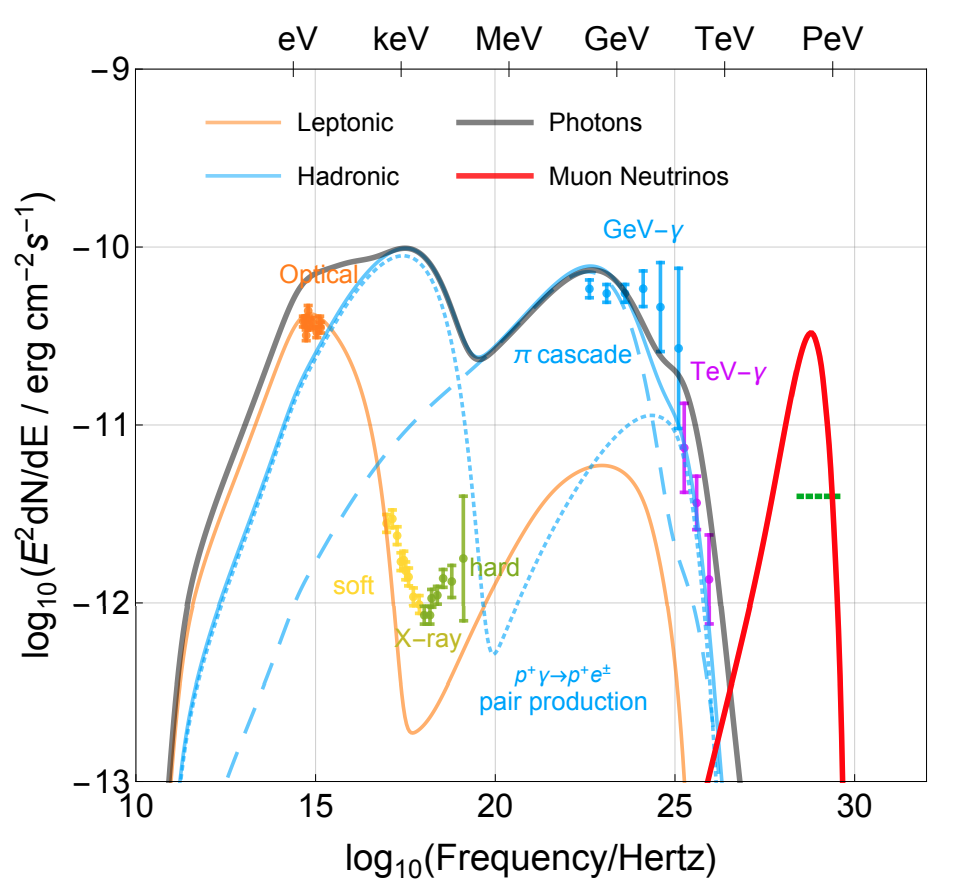}
  \caption{}
  \label{fig:txs_xray_gao}
\end{subfigure}
\caption{Leptohadronic modelling of the Spectral Energy Distribution (SED) of TXS 0506+056 from~\cite{Keivani_2018} \textbf{(a)} and~\cite{Gao_2018} \textbf{(b)}. X-ray data is incompatible with the assumption that gamma-rays in the second hump originate from pion decays, given the neutrino event association is physical.}
\label{fig:txs_xray}
\end{figure}

Leptohadronic emission models for blazars accommodate the possibility of high-energy neutrino production in the jets via $p-\gamma$ (photohadronic) interactions between the accelerated cosmic rays and the ambient or external photon fields. Secondary e$^\pm$ pairs generated by pion decay produce high energy photons via Synchrotron and Compton processes, which in turn combine to produce new pairs which radiate a new generation of photons. These \textit{synchrotron-pair cascades} shift the extreme photon energies down to the X-ray band until the source becomes transparent to the photons and they can escape. Since the origin of the cascade emission and the neutrino flux can be traced back to the $p \pi$ interactions, X-ray data can provide useful constraints on the neutrino flux from the source.~\cite{Keivani_2018} and~\cite{Gao_2018} utlilize the NuSTAR~\cite{Hunter_2010} and SWIFT-XRT~\cite{2005SSRv..120..165B} observations during the 2017 gamma-ray flare of TXS 0506+056 which are almost simultaneous with the detection of IC-170922A~\cite{Aartsen:multimessenger}, to suggest that one-zone models cannot reconcile the X-ray data and the neutrino event from this source assuming that the association of the event with the blazar holds (Fig.~\ref{fig:txs_xray}). Recently,~\cite{Kun_2021} have explored the possibility for PKS 1502+106 and a couple of other blazars, that during a period of enhanced neutrino production, the gamma-ray emission might be subdued (possibly due to absorption within the source) while the X-ray data might have a stronger association with the radio and neutrino emission.

IceCube is a cubic-kilometer neutrino detector installed in ice at the geographic South Pole \cite{Aartsen:2016nxy} between depths of 1450 m and 2450 m. Reconstruction of the direction, energy and flavor of neutrinos relies on the optical detection of Cherenkov radiation emitted by secondary charged particles produced in the interactions of neutrinos with the surrounding ice or the nearby bedrock. The instrumented volume consists of a 3-D array of 5160 Digital Optical Modules (DOMs), each housing an 8-inch photomultiplier tube (PMT) sensitive to Cherenkov photons.

Temporally clustered neutrino emission has been previously reported by IceCube from the direction of TXS 0506+056~\cite{Aartsen:singleflare}. These studies identified the most significant period of emission from the source location over the duration analyzed. However, it is not unlikely for highly variable sources like blazars to have more than one period of slight to moderately increased activity within the continuously increasing length of time for which IceCube has been collecting data. Combining information from multiple flares from a location also provides the benefit of improvement in sensitivity to time-dependent searches.


\section{Analysis overview}

This work will use 10 years of IceCube data to look for 
untriggered time-dependent neutrino emission correlated with Northern sky blazars selected based on their X-ray fluxes. A p-value will be obtained for each blazar by combining the significance of all the neutrino flares detected by performing an unbinned likelihood maximization on the data. These local p-values will be used to perform a statistical study, called the binomial test, to determine and identify if any of the sources in the catalog show a statistically significant emission. The final outcome will be a list of p-values associated with each source, and a binomial p-value for each of the catalogs we test.

The dataset being used for this study spans a period of 10 years from 6th April 2008 to 10th July 2018, including data collected with detector configurations of 40, 59, 79 and 86 strings~\cite{Abbasi:2021datarelease}. It is optimized for all-sky high-energy muon neutrino candidate events, with a background of up-going atmospheric neutrinos and down-going very-high energy atmospheric muons. 

The software used for the analysis (\textit{csky}) fits flares of the box profile, based on an unbinned likelihood maximization (see Section~\ref{sec:mf_method}). The spectral index $\gamma$ and flare duration is fit for each individual flare, while the maximum flare duration to be tested has to be supplied externally.

The cumulative significance of all signal-like flares from a direction will provide a p-value for the source, and these p-values will be used as inputs for the population test using binomial test statistic. The test assumes that p-values are randomly distributed between 0 and 1. Correlations can affect the background expectation of the test. The decorrelation strategy for this analysis involves simulating background trials with a similar configuration and using the same source locations as the actual search, in order that the correlations are present in the background scrambles as well and can be accounted for while comparing with the background distribution. Since the statistical test is more sensitive to a smaller catalog of neutrino-emitting sources, we will test separately for BL Lacs and FSRQs.

\section{A multi-flare search} \label{sec:mf_method}

A neutrino flare can be characterized by an excess of neutrinos or a hardening of the neutrino spectral index observed over a length of time from a particular direction associated with an astrophysical object, that exceeds the background expectation for that period. It can be associated with enhanced neutrino production within the source. Blazars are highly variable sources, with the observed electromagnetic (EM) flares lasting typically for several months. Assuming that neutrino flares show a similar behavior as EM flares, it is reasonable to assume that a source can flare more than once on average within the $\sim$12 year period that IceCube has been taking data. Thus, performing a search for multiple flares over time from each source boosts the significance from each search direction.


In contrast to single-flare searches, which report the single most significant flare for each direction searched, a multi-flare search adds up the significance of all the signal-like flares from a source. It becomes more useful under the hypothesis that most individual flares are moderate to weak and that a source flares more than two times on average. A multi-flare fitting algorithm was developed within IceCube for stacking searches where individual flares can be stacked instead of sources~\cite{luszczak2019method}. The method is described below for a sample of \textit{N} events and \textit{k} source directions:

\begin{enumerate}
    \item The more signal-like events are filtered based on a threshold defined through the ratio of the signal and background PDFs (S/B). This also cuts down on the possible test intervals for flare hypotheses. For this analysis, a threshold of S/B $>$ 2000 is chosen. 
    
    \item Each pair of events in the remaining sub-sample (of $N^{\prime}$ events) presents a possible flare window ($\sim k*N^{\prime}(N^{\prime} - 1)/2$). There can be several overlapping flare windows (see Fig.~\ref{fig:mf_windows})

    \item A test statistic (TS) is calculated for each flare hypothesis, evaluating the likelihood of a real flare in the selected flare window. All events within the window (and not just those above the S/B threshold) contribute to the likelihood.
    
    \begin{equation}
        \mathcal{L}(n_s) = \prod_{i=1}^{N}(\frac{n_s}{N}S_i + (1-\frac{n_s}{N})B_i), \;\;\;\;\;\;\; TS_j|_{\Delta T_j} = -2 log [\frac{\Delta T_{data}}{\Delta T_j} \times \frac{\mathcal{L}(n_s=0))}{\mathcal{L}(n=\hat{n}_s)}]    
    \end{equation}
    
    Here, $\Delta T_j$ refers to the duration of the test window, while $\Delta T$ refers to the total livetime of the sample. The correction factor $\frac{\Delta T_{data}}{\Delta T_j}$ is to adjust for the fact that many more small test windows can be formed than large ones, therefore the small windows are downweighted. 

    \item Only non-negative $TS_j$ are considered, and all the overlapping flares are removed by keeping only the interval with the largest $TS_j$ (Fig.~\ref{fig:mf_overlap}). This way individual events don't end up contributing to more than one flare. A global TS can be obtained by summing all the remaining $TS_j$. A '\textit{multi-flare TS}' for each source can also be obtained in the same way by summing all the individual flares from the source direction.
    
    \begin{equation}
    \tilde{TS} = \sum_{TS_j>0}TS_j  \;\;and \;\; \tilde{n_s} = \sum_{TS_j>0}n_{sj} 
    \label{eq:ts_ns}
    \end{equation}

    \begin{figure}[h]
    \centering
    \begin{subfigure}{.4\textwidth}
      \centering
      \includegraphics[width=\linewidth]{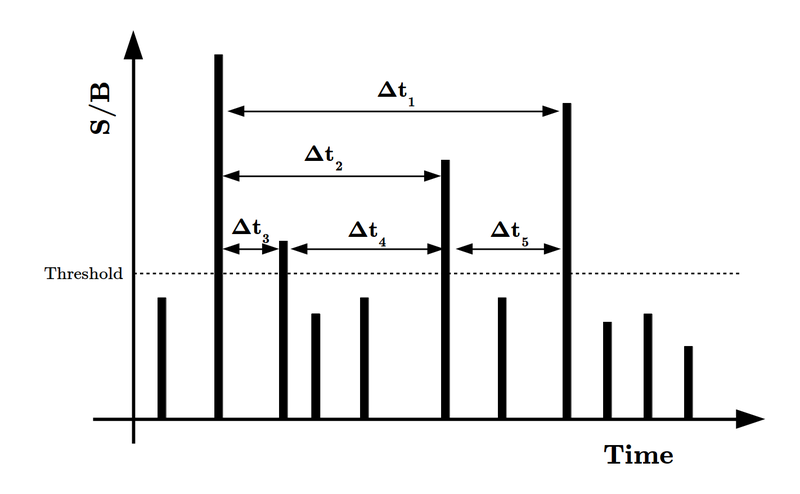}
      \caption{}
      \label{fig:mf_windows}
    \end{subfigure}%
     \begin{subfigure}{.55\textwidth}
      \centering
      \vspace{1.35cm}\includegraphics[width=\linewidth]{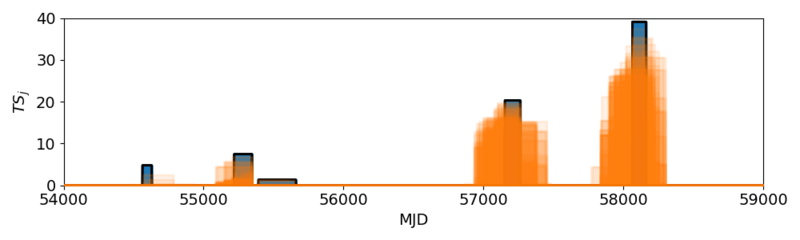}
      \caption{}
      \label{fig:mf_overlap}
    \end{subfigure}
    \caption{Flare window construction using seed events \textbf{(a)}, and a demonstrative figure underlining the removal of overlapping flares \textbf{(b)}. The orange flare candidates end up getting removed in favor of the blue flare candidates that are of higher significance.}
    \label{fig:mf}
    \end{figure}

\end{enumerate}

The above method returns some global parameters: TS, $n_s$ (sum of the number of signal events above the average background expectation, detected during each flare duration), average spectral index $\gamma$, number of flare hypotheses tested, as well as the fit parameters for the individual flares, including $TS_j$, $n_{sj}$, $\gamma$ and duration of flare. The multi-flare TS for each source can be converted to a corresponding (one sided) pre-trial p-value by comparing with the TS distribution from several background trials using simulated data. The simulated data (called scrambled data) is generated by randomizing the right ascension (RA) of the events in real data so that the physical properties of the data are preserved.



\section{Binomial Test}

A catalog search has the possibility of revealing a few signal sources spread among a background of sources. While the sources may not be individually significant, it is nonetheless important to understand if the whole catalog is compatible with the background-only hypothesis. The binomial test is a statistical method to determine if a subset of values in a given sample exceed the expectation from a random distribution of the sample size. Given an ordered set of p-values, it can determine if a subset of these values are statistically significant or not.

The binomial probability $P(k)$ of finding $k$ or more sources with p-values equal to or lower than the local p-value $p_k$, in an ordered list of $N$ p-values is given by: 

\begin{equation}
    P(k)=\sum_{m=k}^N \binom{N}{m}p_k^m(1-p_k)^{N-m}
\end{equation}

It returns a binomial p-value ($P(k)$) for each source and a $k$ corresponding to the best binomial p-value. The binomial p-value is an indicator of the individual significance of time-dependent emission from the source. $k$ identifies the number of sources in the catalog with a statistically significant emission. 

\begin{figure}[h]
\centering
\includegraphics[width=0.4\textwidth]{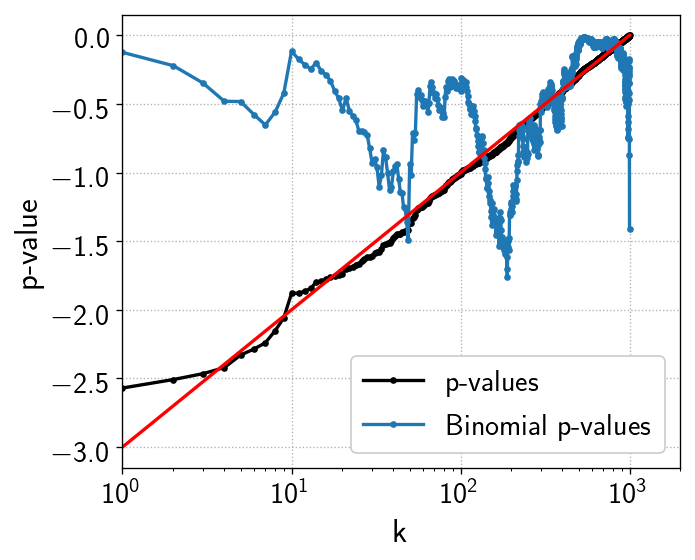}
\caption{Binomial probability distribution for a single binomial test with 1000 simulated p-values. The best binomial probability $B_{best} = 0.017$ is achieved for a $k$ of 190.}
\label{fig:bt_test}
 \end{figure}
 
\begin{figure}[h]
\centering
\begin{subfigure}{.4\textwidth}
  \centering
  \includegraphics[width=\linewidth]{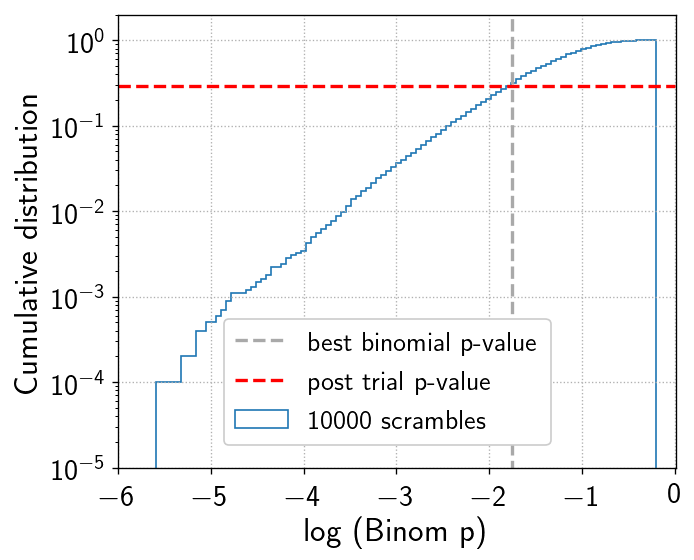}
  \caption{}
  \label{fig:bt_posttrial}
\end{subfigure}%
\hspace{1cm}
 \begin{subfigure}{.4\textwidth}
  \centering
  \includegraphics[width=\linewidth]{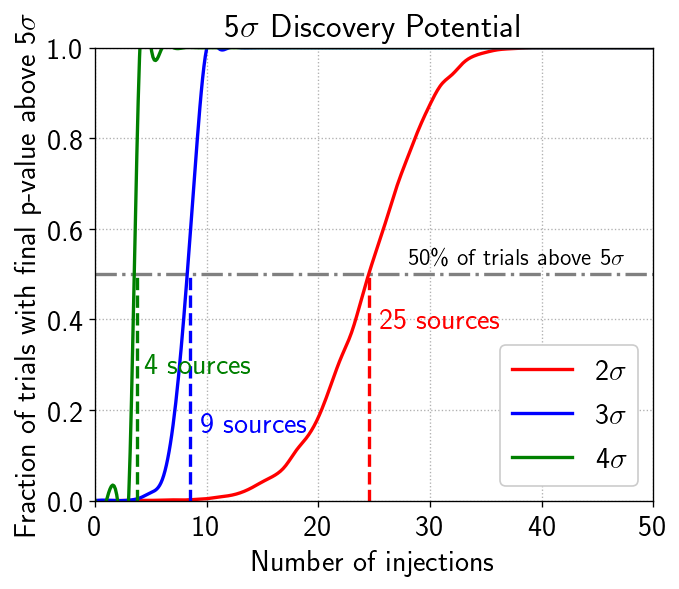}
  \caption{}
  \label{fig:bt_flaresig}
\end{subfigure}
\caption{Post-trial p-value calculation by comparing with 10000 binomial tests performed on simulated data \textbf{(a)}, and the
number of individual sources with $2\sigma$, $3\sigma$ and $4\sigma$ multiflare significance required to obtain a final discovery potential of $5\sigma$ in 50\% of pseudo-experiments \textbf{(b)}.}
\label{fig:bt_sig}
\end{figure}

An example of the output of a single binomial test, performed on simulated p-values is shown in Fig.~\ref{fig:bt_test}. The final significance is derived by performing the binomial test many times over scrambled data, to trial correct for looking at multiple locations in the sky. Fig.~\ref{fig:bt_posttrial} shows the calculation of post trial p-value by comparing with the results from multiple trials with simulated data and Fig.~\ref{fig:bt_flaresig} shows the number of individual sources with $2\sigma$, $3\sigma$ and $4\sigma$ multi-flare significance required to obtain a final discovery potential of $5\sigma$ from the analysis.

\section{X-ray selected blazar catalog}

While gamma-ray catalogs of Active Galactic Nuclei (AGN) have existed for a while now and are among the most detailed catalogs for this class of sources ~\cite{Ackermann_2015}, few attempts have been made to compile X-ray catalogs of blazars. Although an exclusive catalog for X-ray selected BL Lacs exists, the XRAYSELBLL~\cite{2013AJ....145...31K}, it contains only $\sim$300 sources and comprises solely of BL Lac type blazars. Multi-frequency catalogs, however, do exist for blazars. Among the most recent and comprehensive of these is the $5^{th}$ edition of RomaBZCAT~\cite{2015Ap&SS.357...75M}. It contains information on 3561 blazar AGNs (BL Lacs, FSRQs and blazars of uncertain type). Apart from positions and redshifts wherever available, the data is available for radio fluxes in the 1.4 GHz band from NVSS, microwave band (143 GHz) from the Planck survey, the R-band magnitude, soft X-ray flux in (0.1 - 2.4 keV) from ROSAT, and gamma-ray flux in (1-100 GeV) from Fermi-LAT. The catalog has a comprehensive sky coverage, with sources covering the entire Northern and Southern sky. XRAYSELBLL has a $> 60\%$ overlap with RomaBZCat.

\begin{figure}[h]
\centering
\begin{subfigure}{.52\textwidth}
  \centering
  \includegraphics[width=\linewidth]{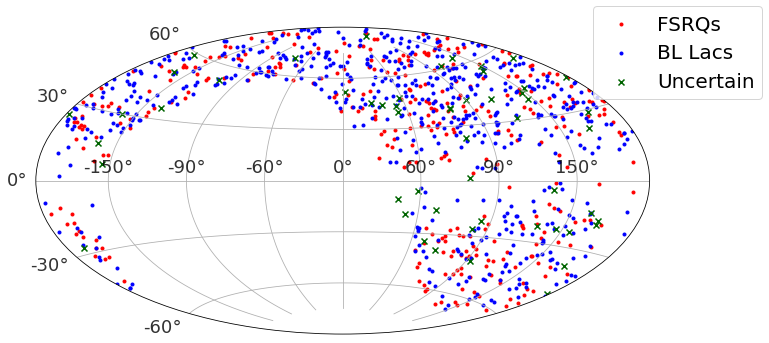}
  \caption{}
  \label{fig:cat_skygal}
\end{subfigure}%
\hspace{1cm}
 \begin{subfigure}{.4\textwidth}
  \centering
  \includegraphics[width=\linewidth]{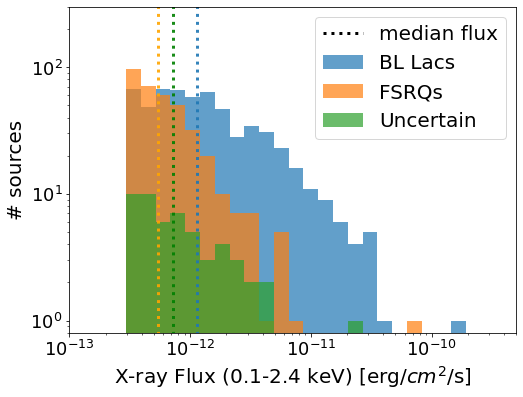}
  \caption{}
  \label{fig:cat_fluxes}
\end{subfigure}
\caption{Skymap of the 1000 sources with the highest X-ray fluxes in the declination ($-5^{\circ}$, $+85^{\circ}$) from RomaBZCat shown in Galactic coordinates \textbf{(a)}, and the distribution of the X-ray fluxes for these 1000 sources \textbf{(b)}. The dotted lines indicate the median X-ray flux for each category.}
\label{fig:cat_sky}
\end{figure}

For the purpose of this analysis, we order the the sources in the catalog by their X-ray fluxes and keep only the top 1000 blazars in the Northern hemisphere ($-5^{\circ}$, $+85^{\circ}$). The distribution of these 1000 sources in the sky and their X-ray fluxes can be seen in Fig.~\ref{fig:cat_sky}.

The down-selected catalog contains 586 BL Lacs, 361 FSRQs and 53 uncertain type blazars. The minimum and maximum X-ray fluxes are $3.1 \times 10^{-13}$ and $1.8 \times 10^{-10}$ erg/cm$^2$/s respectively. Overlap between our catalog and a previous study that looked at the neutrino flares from top 1000 Northern sky blazars of Fermi 3LAC catalog~\cite{Ackermann_2015} is $\sim$44\%. 

\section{Per source sensitivity}

\begin{figure}[h]
\centering
\begin{subfigure}{.42\textwidth}
  \centering
  \includegraphics[width=\linewidth]{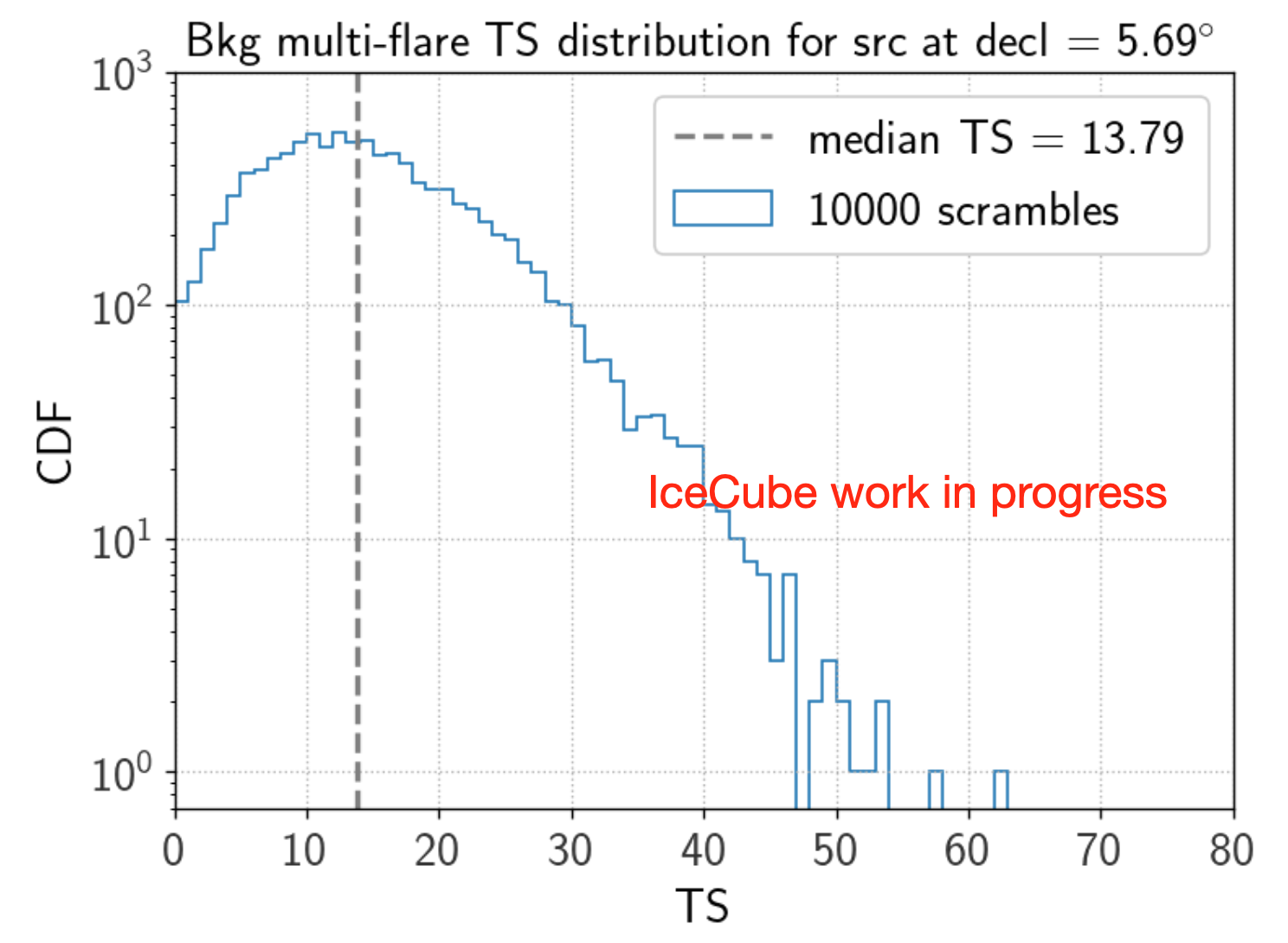}
  \caption{}
  \label{fig:persrc_bgTS}
\end{subfigure}%
\hspace{1cm}
 \begin{subfigure}{.4\textwidth}
  \centering
  \includegraphics[width=\linewidth]{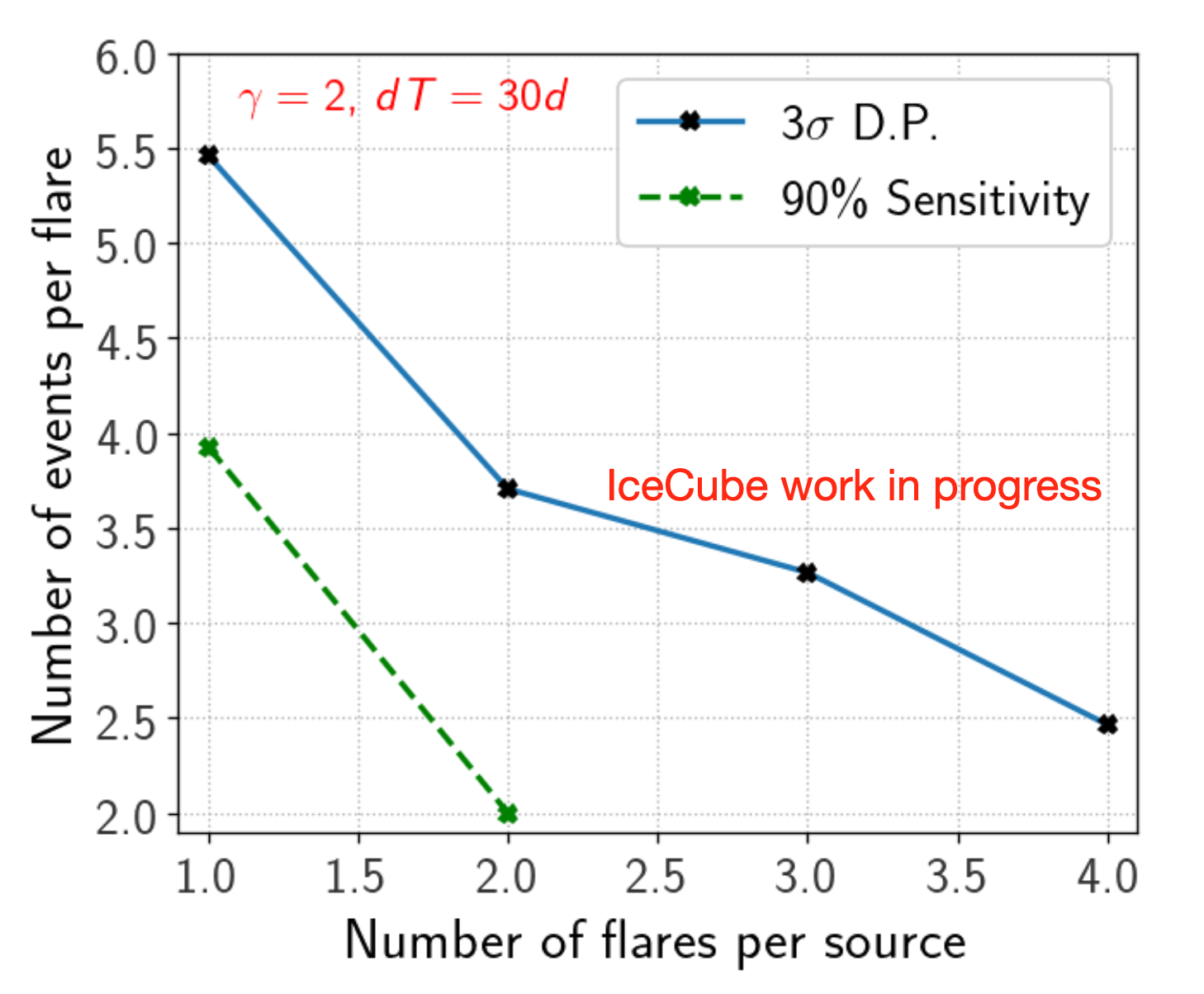}
  \caption{}
  \label{fig:persrc_3sig_DP}
\end{subfigure}
\caption{Background multi-flare TS distribution for a source at the declination of TXS 0506+056 (5.69$^{\circ}$) \textbf{(a)}, and   the strength of individual flares from a single source (at the declination of TXS 0506+056) required to obtain a final D.P. of 3-sigma and a 90\% C.L. sensitivity \textbf{(b)}. The 5-sigma D.P. is not shown here due to limited statistics but will be available in future tests.}
\label{fig:per_src}
\end{figure}

Sensitivity studies for a single source were performed by injecting signal flares of varying strength at arbitrary times at the location of TXS 0506+056 (declination = 5.69$^{\circ}$). The flare strength was varied by changing the number of flares injected as well as the number of events injected per flare. The spectral index ($\gamma = 2$) and the flare duration (dT = 30 days) were kept constant for all the injections. These were compared with background trials to obtain the sensitivity of the analysis to a single source. The number of events per flare vs. the number of flares required to obtain a 90\% C.L. sensitivity and a 3$\sigma$ discovery potential (D.P.) with an S/B threshold of 2000 are shown in Fig.~\ref{fig:persrc_3sig_DP}.  

\section{Summary and Outlook}

We have presented here an outline for our intended search for time-clustered neutrino emission in IceCube data from blazars selected on the basis of their soft X-ray fluxes (0.1 - 2.4 keV). We aim to test the hypothesis that X-ray bright blazars can be potential sources of high-energy astrophysical neutrinos. The importance of X-ray data when trying to constrain the neutrino emission of blazars is explained through the recent attempts at leptohadronic modelling of the SED of TXS 0506+056. Additionally, this analysis has the advantage of accumulating information from flares observed over 10 years of IceCube neutrino data. A brief description of the multi-flare fit method and a population test through the binomial test statistic is also provided. The catalog of sources is derived from RomaBZCat by selecting 1000 blazars with the highest X-ray flux in the Northern sky. The dataset and other parameters to be used for the analysis have been defined and background trials are currently under progress. Sensitivity of the analysis to a single source is presented here, while the full catalog sensitivity and a final p-value for each source signifying their time-dependent neutrino emission potential will be published soon.



\bibliographystyle{ICRC}
\bibliography{references}

%
%
%


\end{document}